\documentclass{iau} 
\usepackage{hyperref}
\usepackage{graphicx}
\usepackage{subfig}
\usepackage{xcolor}
\usepackage{xparse}

\ExplSyntaxOn

\keys_define:nn { miguel/label }
 {
  label   .tl_set:N = \l_miguel_label_tl,
  unknown .code:n   = \clist_put_right:Nx \l_miguel_label_clist
                       { \l_keys_key_tl = \exp_not:n { #1 } }
 }
\clist_new:N \l_miguel_label_clist
\box_new:N \l_miguel_label_box
\box_new:N \l_miguel_label_image_box

\NewDocumentCommand{\xincludegraphics}{O{}m}
 {
  \tl_clear:N \l_miguel_label_tl
  \clist_clear:N \l_miguel_label_clist
  \keys_set:nn { miguel/label } { #1 }
  \tl_if_empty:NTF \l_miguel_label_tl
   {
    \miguel_includegraphics:Vn \l_miguel_label_clist { #2 }
   }
   {
    \hbox_set:Nn \l_miguel_label_image_box
     {
      \miguel_includegraphics:Vn \l_miguel_label_clist { #2 }
     }
    \hbox_set:Nn \l_miguel_label_box
     {
      \skip_horizontal:n { 55pt }
      \fcolorbox{white}{white}{\footnotesize \tl_use:N \l_miguel_label_tl}
     }
    \leavevmode
    \box_use:N \l_miguel_label_image_box
    \skip_horizontal:n { -\box_wd:N \l_miguel_label_image_box }
    \hbox_overlap_right:n
     {
      \box_move_up:nn
       {
        \box_ht:N \l_miguel_label_image_box - 
        \box_ht:N \l_miguel_label_box - 3pt
       }
       { \box_use_drop:N \l_miguel_label_box }
     }
    \skip_horizontal:n { \box_wd:N \l_miguel_label_image_box }
   }
 }

\cs_new_protected:Nn \miguel_includegraphics:nn
 {
  \includegraphics[#1]{#2}
 }
\cs_generate_variant:Nn \miguel_includegraphics:nn { V }
\usepackage{amsmath}
\usepackage{amssymb}

\ExplSyntaxOff

\title[Magnetar Spectral Lines with Future Telescopes] 
{Probing Magnetars Using Spectral Lines with Future Telescopes
}

\author[Demet K{\i}rm{\i}z{\i}bayrak \& Jeremy Heyl]   
{Demet K{\i}rm{\i}z{\i}bayrak$^1$
 \and Jeremy Heyl$^1$}

\affiliation{$^1$Department of Physics \& Astronomy, University of British Columbia, Vancouver, BC V6T 1Z1, Canada; e-mail: {\tt demet@phas.ubc.ca}, {\tt heyl@phas.ubc.ca}  }

\pubyear{2021}
\volume{363}  
\jname{Neutron Star Astrophysics at the Crossroads: Magnetars and the Multimessenger Revolution}
\editors{E. Troja \& M. Baring, eds.}
\begin{document}

\maketitle

\begin{abstract}
We present our findings on magnetar spectral line analysis in the context of upcoming high resolution, high effective area, high throughput X-ray telescopes for two cases: persistent magnetar emission and magnetar bursts. For magnetars in quiescence, we present our preliminary work on modelling for phase-resolved emission. Our results reveal the necessity of constraining line depth and width concurrently with line energy to conclusively determine hotspot emission and corresponding magnetic field geometry. We then present the results of our simulations using effective area and response of various current and upcoming X-ray telescopes for magnetar spectral line detection and expand on the exciting opportunities upcoming telescopes provide to probe quiescent and burst emission region geometry and propagation in the extreme magnetic field of a magnetar.

\keywords{stars: neutron – stars: magnetic fields - X-rays: bursts - X-rays: general - techniques: spectroscopic - telescopes }
\end{abstract}

\firstsection 
\section{Introduction}

Magnetar spectral lines are thought to result from proton cyclotron scattering in the magnetosphere, where photons emerging from a hotspot on the surface are scattered at  the cyclotron frequency, resulting in absorption features in the spectrum. However, relativistic effects due to the magnetar's gravitational field, as well as changes in the extreme magnetic field near the emission region highly broaden line features, making them much harder to detect. 

Studying magnetar spectral lines promises insights into many unknown features of magnetars and physics in strong magnetic fields. First, if interpreted in the context of the magnetar model, spectral lines yield an estimate of the magnetar's magnetic and gravitational fields near the crust and the size of the emission region. Moreover, spectral lines offer a unique tool to resolve the geometry of the emission region as well as that of the field lines near the surface. Understanding such geometry would enable astronomers to identify exact causes of the quiescent and burst emission. 

\section{Modelling Phase-Resolved Spectroscopy}

\begin{figure}[htp]
 \centering
\includegraphics[trim={4cm 1cm 7cm 2cm},clip,width=4cm]{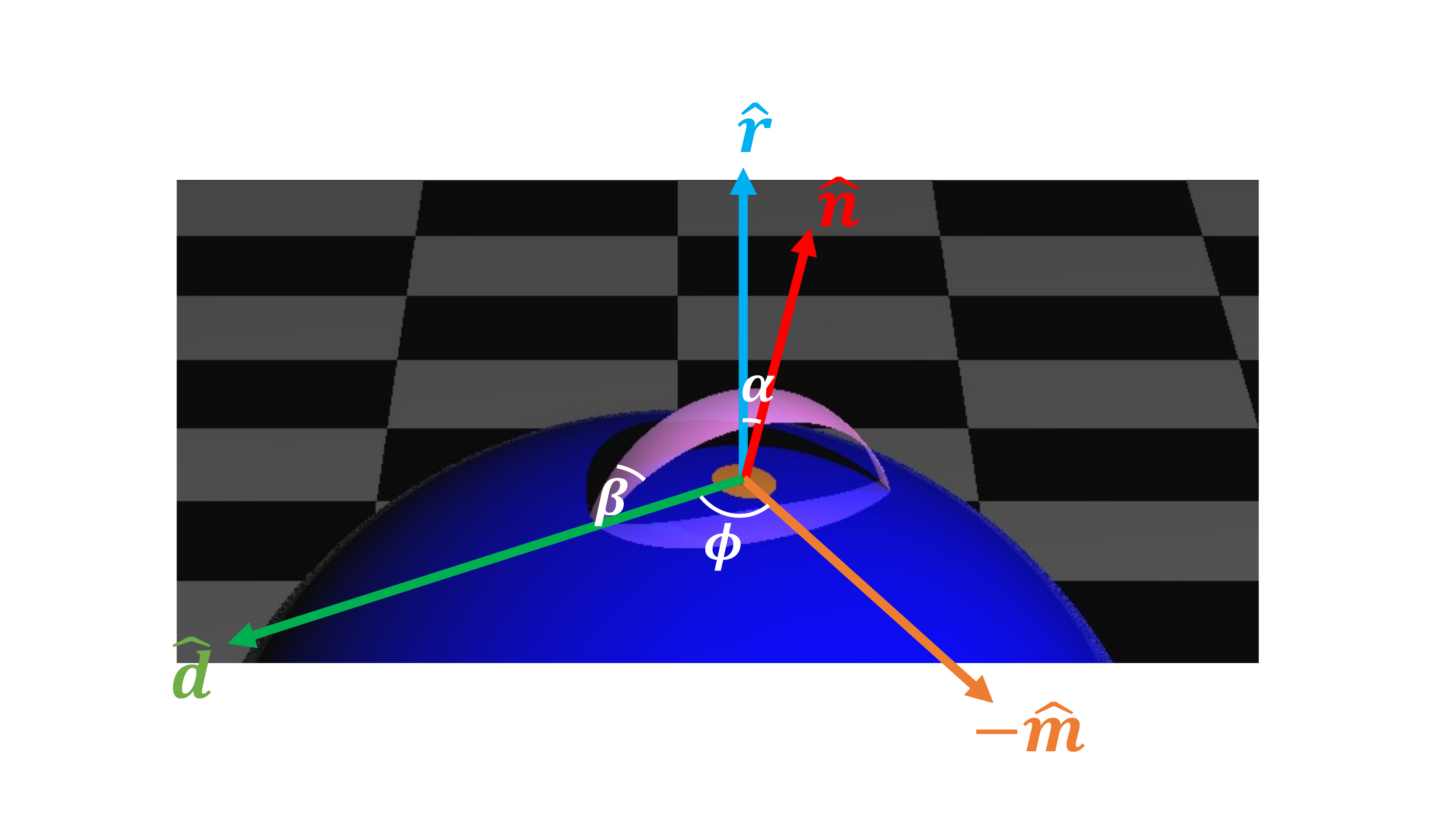}
\centering
\includegraphics[trim={0 -2.2cm 0 0 },clip,width=3cm]{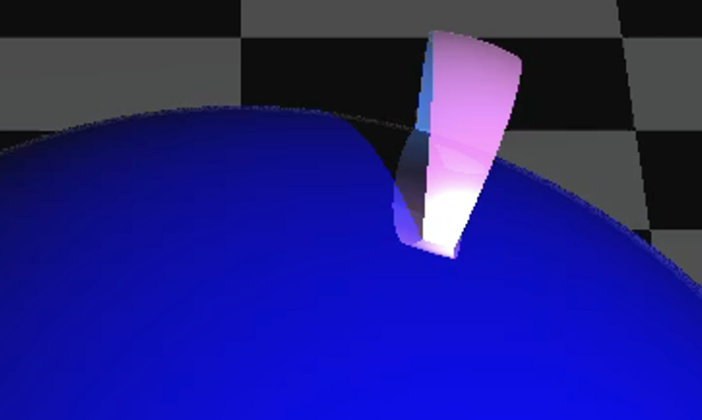}
\includegraphics[trim={0 -2.2cm 0 0.45cm },clip,width=3cm]{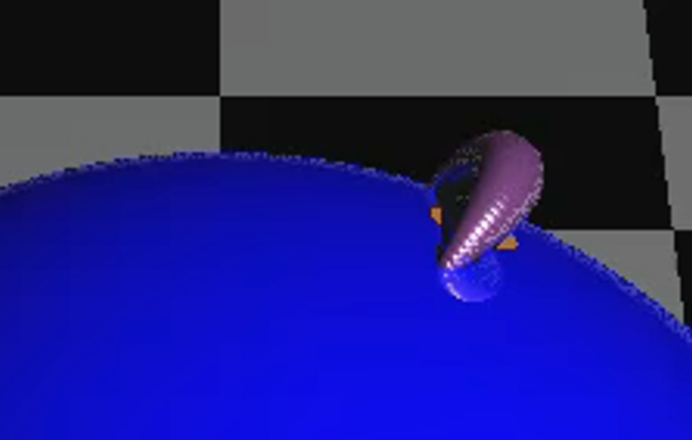} 
\caption{\textit{Left:} Visualization of magnetar emission geometry model with a hotspot (orange) and loop of magnetic field lines carrying along a current (pink). $\hat{r}, \hat{n}, \hat{d}$ and $\hat{m}$ represent the rotation axis, spot normal, loop diameter and spot meridian respectively. \textit{Middle:} Visualization of a ribbon-like loop. \textit{Right:} Visualization of a tube-like loop.
\label{fig:mag_geo}}
\end{figure}

In the proton cyclotron model, changes in spectral lines throughout a magnetar's rotation phase can be explained as resulting from the geometry of the magnetic field lines surrounding the emitting region. Here, we use the cyclotron absorption model 
\begin{equation}
    F(E) = F_0 \ \exp{\Big[-D \dfrac{(W E / E_c)^2}{(E-E_c)^2+W^2}\Big]}
    \label{equation:cycl_model}
\end{equation}
from \cite[Tiengo et. al. (2013)]{tiengo2013} and \cite[Makishima et. al. (1990)]{makishima1990} where $D$ and $W$ are line depth and width, $F_0$ is the level of the continuum, $F(E)$ is the flux at the given energy, and $E_c$ is the cyclotron energy. A key observation was discovered by \cite[Tiengo et. al. (2013)]{tiengo2013} where absorption lines were observed to vary in energy, width and depth as the magnetar SGR 0418+5729 rotates during its persistent emission. To model this behaviour, we assume a solar prominence-like scenario with a hotspot on the surface emitting photons that travel through a current-loaded loop of plasma along magnetic field lines as in \cite[Tiengo et. al. (2013)]{tiengo2013}. As the magnetar rotates, photons emerging from the hotspot travel through different locations on the transverse loop to reach our line of sight. The location on the loop where photons traverse affects the line energy, as photons meeting with the narrower base of the loop encounter a stronger field, resulting in a higher line energy. On the left panel of Figure \ref{fig:mag_geo}, we illustrate this model and several other parameters expected to affect line shape: hotspot location ($\alpha$), loop orientation ($\phi$) and loop transverse angle ($\beta$). 

\begin{figure*}[htp]
\centering
\includegraphics[trim={0.2cm 0 0.3cm 0},clip,scale=0.3]{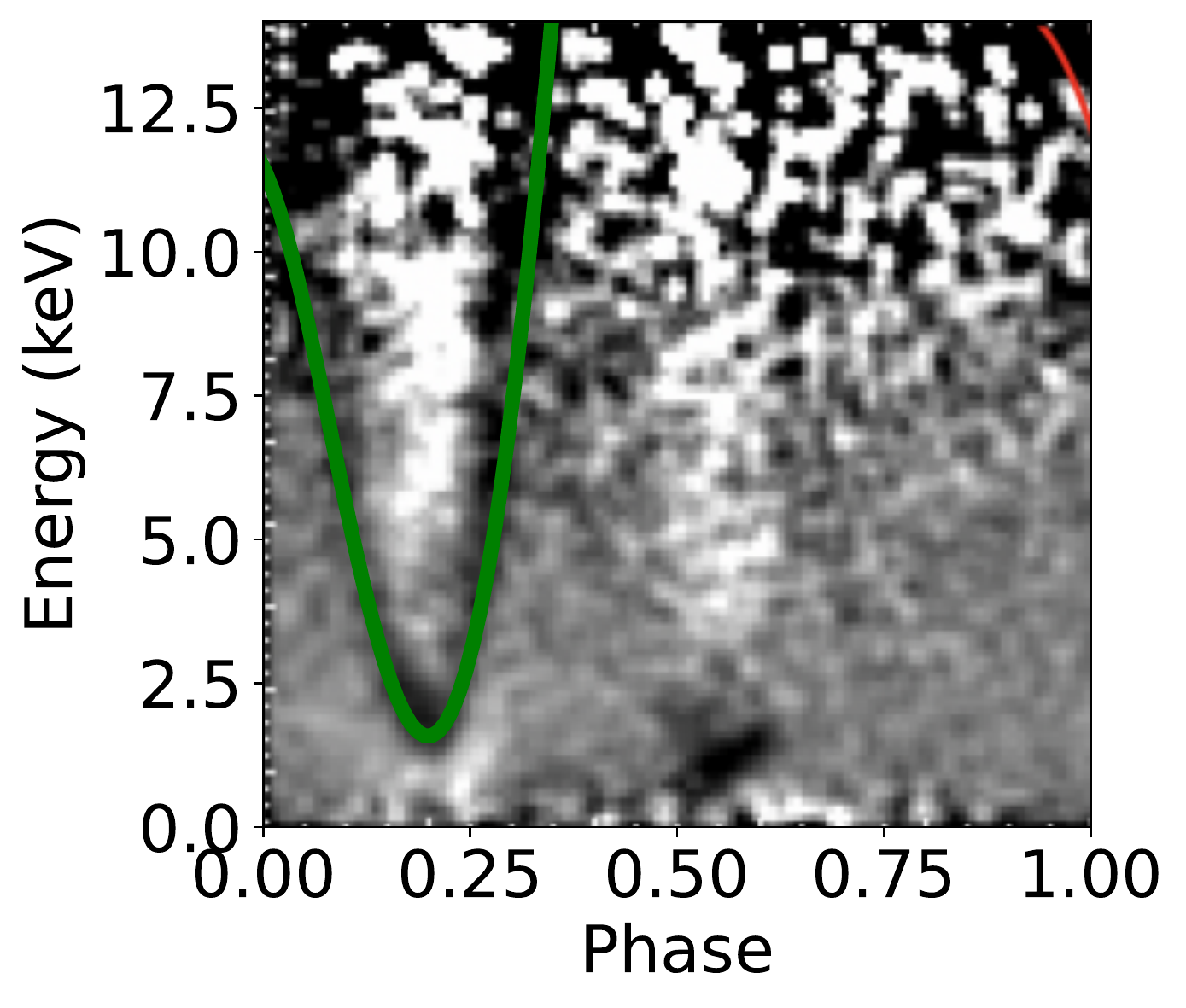}
\includegraphics[trim={0.2cm 0 0.3cm 0},clip,scale=0.3]{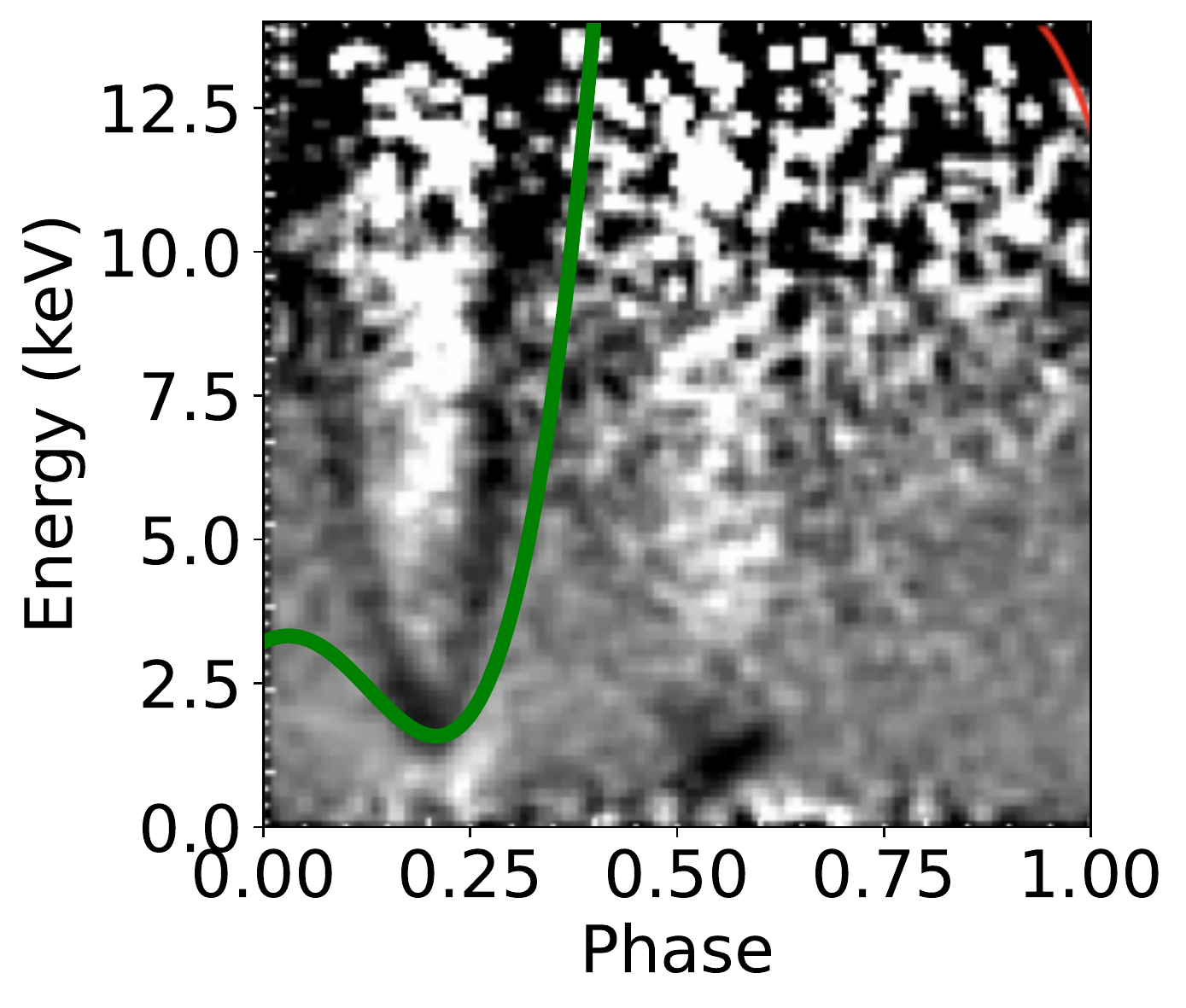}
\includegraphics[trim={0.2cm 0 0.3cm 0},clip,scale=0.3]{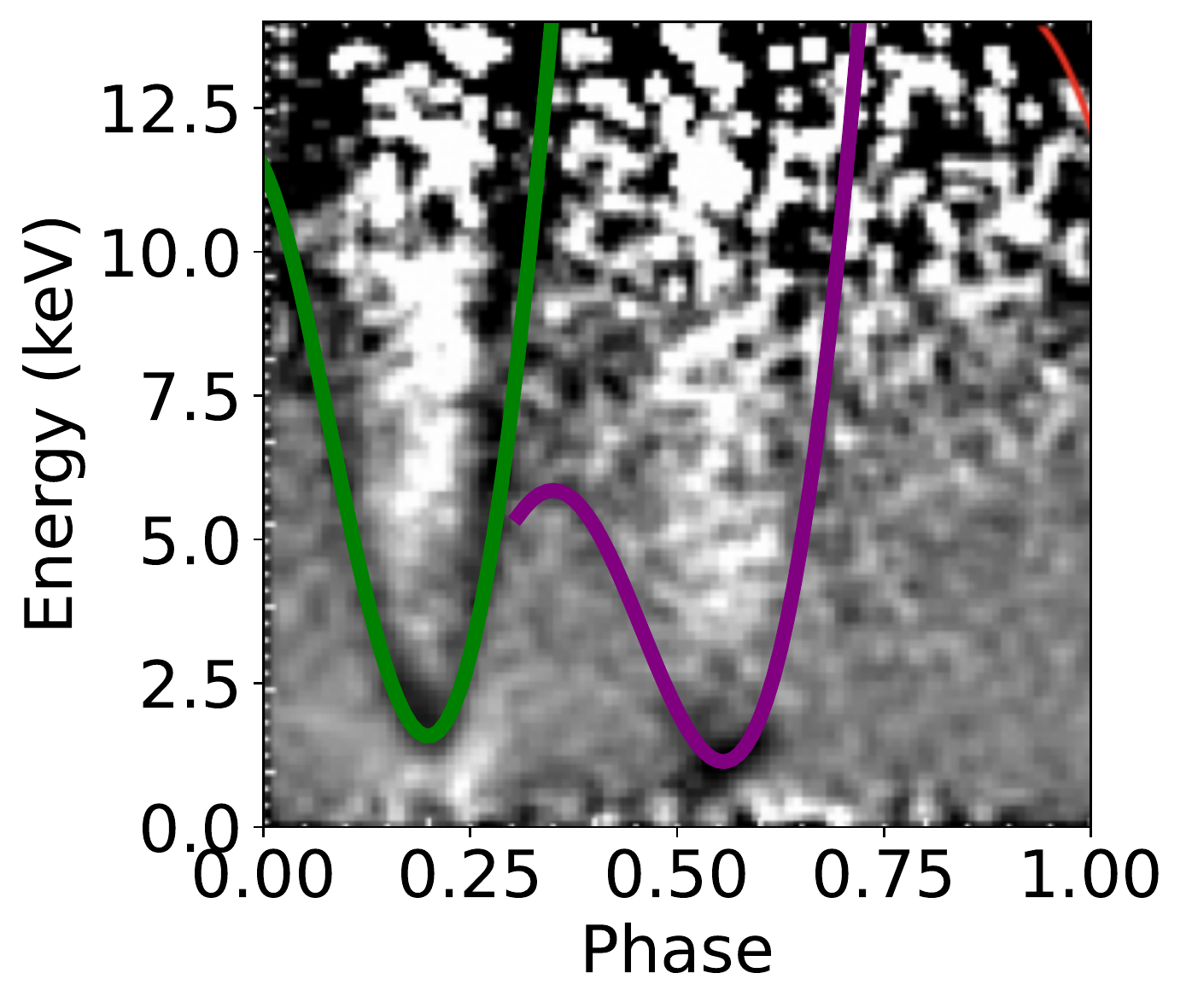}
\includegraphics[trim ={0 -4.5cm 0 0},height= 3.6cm]{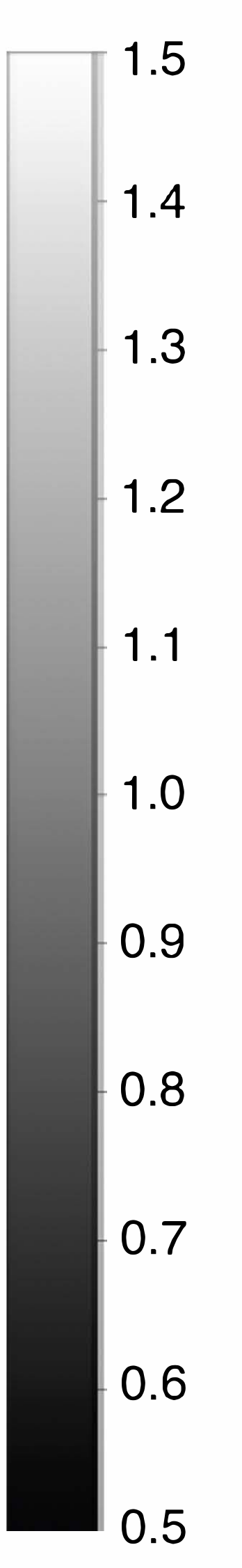}
\caption{Phase Resolved Spectroscopy of SGR 0418+5729 from \cite[Tiengo et. al. (2013)]{tiengo2013} and model discussed in text. Darker regions indicate a lack of counts and green lines depict our model with a hotspot at \textit{Left:} $20^{\circ}$ and \textit{Middle:} $50^{\circ}$ from the rotation axis. \textit{Right:} Model with a second hotspot (purple) with larger width gradient at $37^{\circ}$ from the rotation axis. 
\label{fig:phaseres_modelfit}}
\end{figure*}

We plot the resulting model using parameters provided in \cite[Tiengo et. al. (2013)]{tiengo2013} on top of the phase-resolved spectrum of SGR 0418+5729 observed with XMM-Newton EPIC-pn on the left panel of Figure \ref{fig:phaseres_modelfit}. In the phase-resolved spectrum, darker regions indicating an absorption line trace a V-shape as the magnetar rotates between phases $\sim$0$-$0.4. The model is able to obtain this with $\alpha = 20^{\circ}$, $\phi= 90^{\circ}$ and the line of sight at $70^{\circ}$. To demonstrate the sensitivity for different geometries, we plot the same model with a hotspot closer to the magnetar's equator ($\alpha =50^{\circ}$) keeping all remaining parameters constant on the middle panel of Figure \ref{fig:phaseres_modelfit}. In this case, the model fails to trace the observed line energy, exhibiting our ability to identify hotspot location, loop orientation and line of sight using spectral line energy.

We note that a hint of a second broader line with a lower dip is present between phases $\sim$0.4$-$0.7. We are not able to explain this feature using a single hotspot model. However, with a second hotspot at $\alpha =37^{\circ}$ from the rotation axis, and a larger field gradient ($f = 7.23 \times 10^{15} \ G$), we obtain the purple line on the right panel of Figure \ref{fig:phaseres_modelfit}. Yet, since the detection is weak, it is not possible to conclusively determine the hotspot parameters or rule out any other scenarios. More data, especially with future higher effective area telescopes could allow to probe such features to understand hotspot origins. 

\begin{figure}[htp]
\centering
\includegraphics[trim={0.2cm 0 0 0},clip,width=0.27\textwidth]{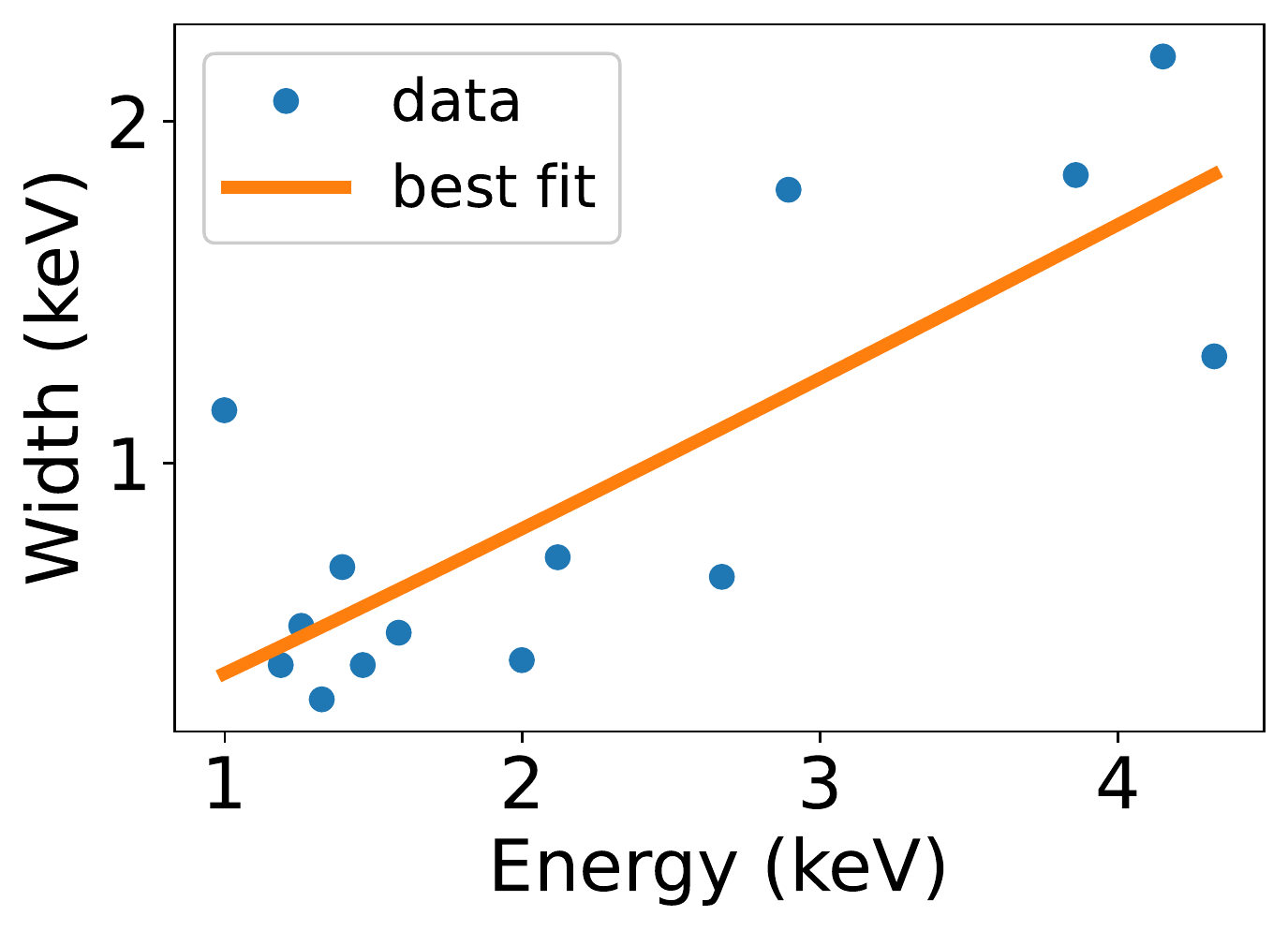}
\includegraphics[trim={0.2cm 0 0 0},clip,width=0.3\textwidth]{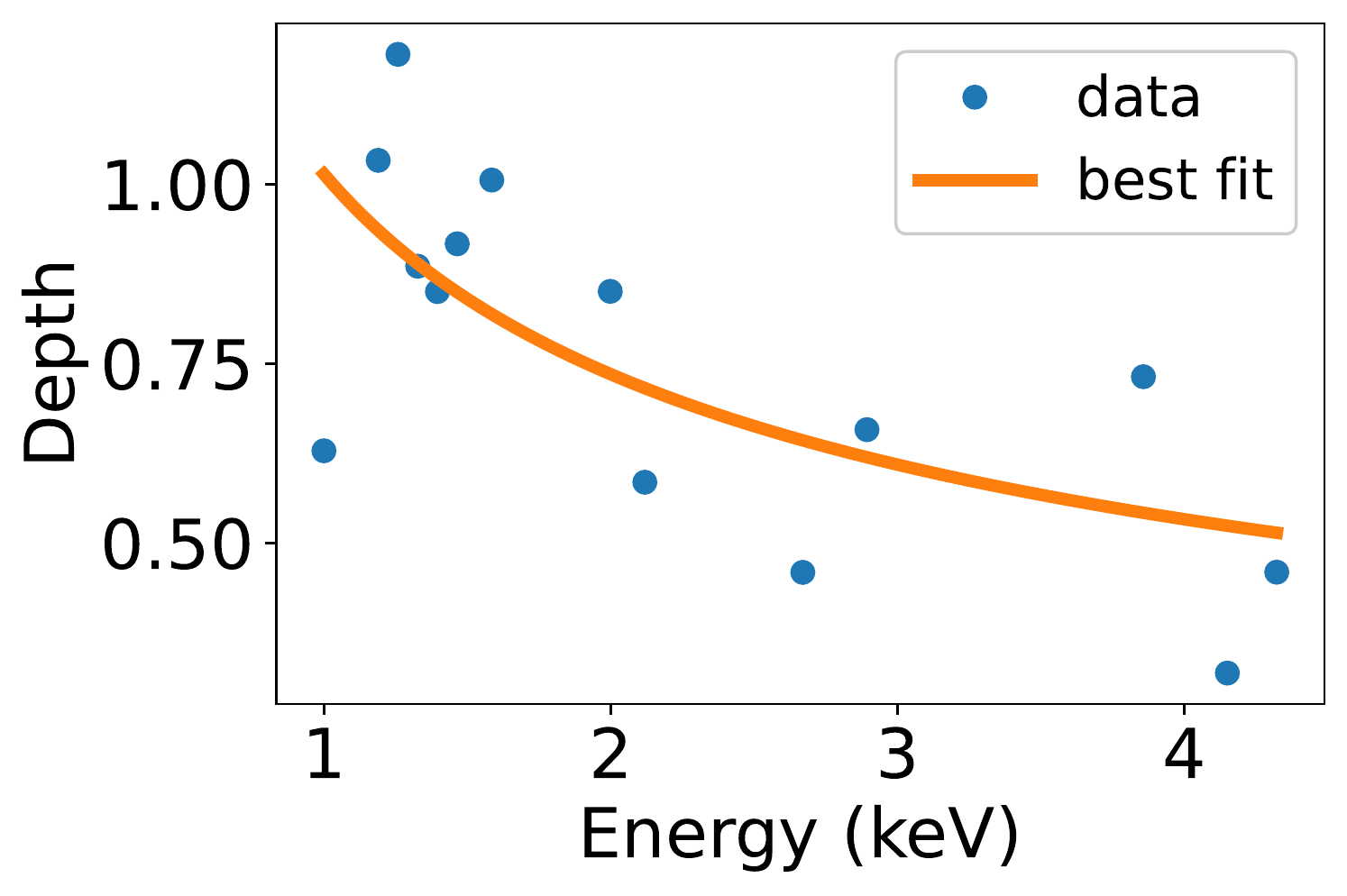}
\includegraphics[trim={0 -0.9cm 0 0cm },clip,width=0.3\textwidth]{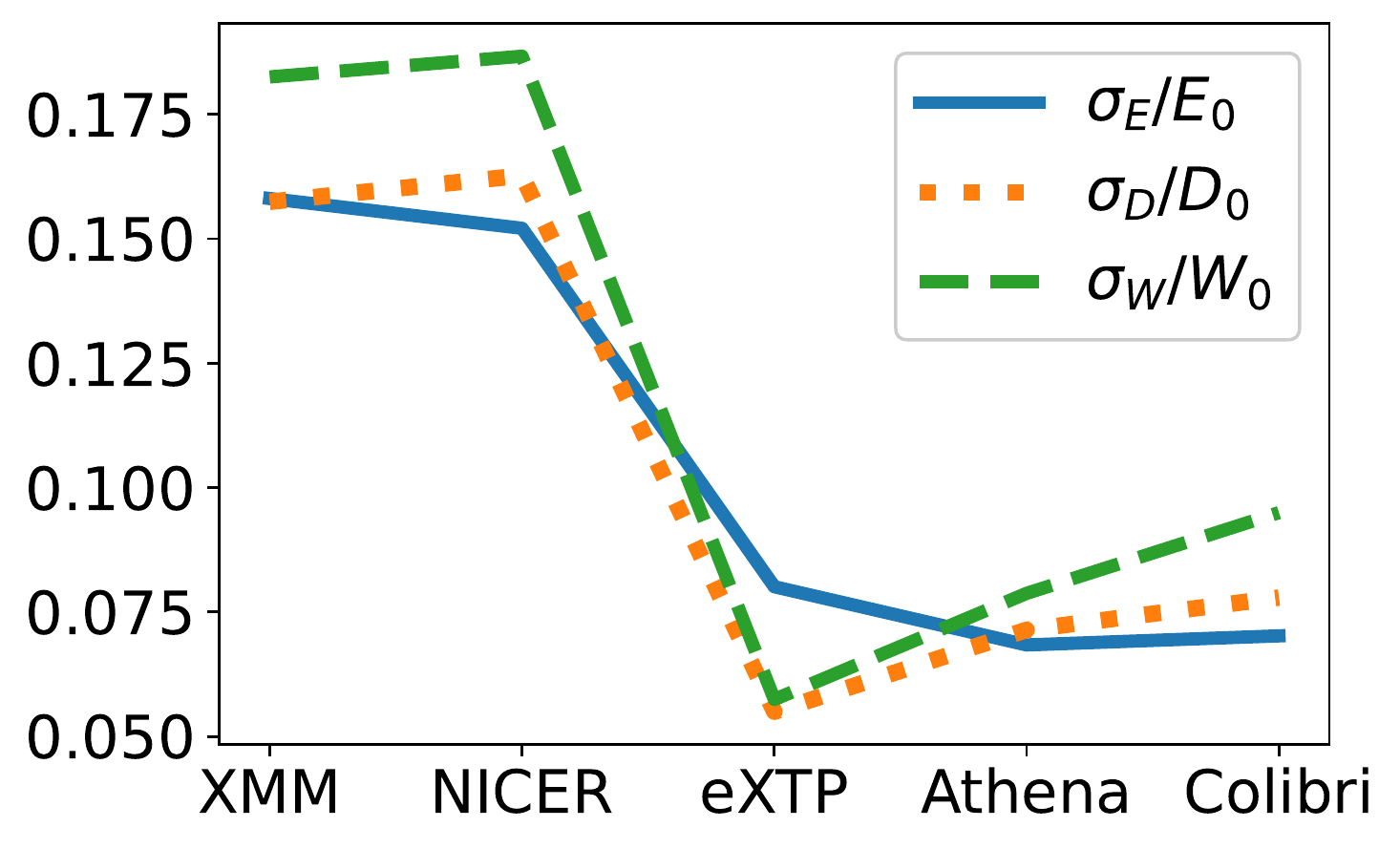}
\caption{\textit{Left:} Width (keV) vs. Energy (keV), \textit{Middle:} Depth vs. Energy (keV) of all lines detected throughout the phase of SGR 0418+5729 in \cite[Tiengo et. al. 2013]{tiengo2013}. Orange lines indicate best power-law fits with $\textrm{Width} \sim \textrm{Energy}$ and $\textrm{Depth} \sim \textrm{Energy}^{-0.5}$ within error bars, indicating a scenario closer to a tube-like loop in Figure \ref{fig:mag_geo}. \textit{Right:} Average 1$\sigma$ Energy (blue), Depth (orange) and Width (green) errors per fitted value for all SGR 0418+5729 spectral lines between between phases 0$-$0.3 in \cite[Tiengo et. al. 2013]{tiengo2013} vs. simulated detectors.
\label{fig:edw}}
\end{figure}

Although line width and depth are also observed to vary as SGR 0418+5729 rotates, this behaviour was not yet studied. To demonstrate what these features can tell us about magnetic field geometry, we consider two cases: 1) A ribbon-like transverse loop 2) A tube-like loop with a certain thickness as well as a width (illustrated in Figure \ref{fig:mag_geo}). Here, the line equivalent width can be expressed as
\begin{equation} 
\Gamma (E) = \int (1- F(E)/F_0) \,dE
\end{equation}
where $\Gamma$ is the equivalent width, $F_0$ the continuum level, and $F(E)$ the line flux across the energy of interest. Line equivalent width depends on the number of field lines photons encounter on the loop, and the magnetic field scales with the field line density, yielding:
\begin{equation}
    \Gamma \propto L^{-1} \ \textrm{(ribbon \& tube)} \hspace{1cm} B \propto L^{-1} \ \textrm{(ribbon)} \hspace{1cm}  B\propto L^{-2} \ \textrm{(tube)}
\label{equation:eqwidth}
\end{equation}
where $B$ is the magnetic field and $L$ is the length scale of the loop. Considering all spectral lines detected for SGR 0418+5279 from \cite[Tiengo et. al. 2013]{tiengo2013}, our best power-law fits show that line energy is linear with line width within error bars ($W \propto A \times E^{\alpha} $ where $A$ is the amplitude and $\alpha = 1.07 \pm 0.26$, Figure \ref{fig:edw}). Assuming that this is correct, and using Equation \ref{equation:eqwidth}, and $\Gamma \propto DW$ we get:
\begin{equation}
    D \propto E^0 \  \textrm{(ribbon)} \hspace{1cm} D \propto E^{-0.5} \ \textrm{(tube)}
\end{equation}
where $D$ is the line depth, and $E$ is the line energy. For SGR 0418+5279 line depth vs. energy (Figure \ref{fig:edw}), we obtain a best fit power of 0.5 within error bars (($D \propto A \times E^{\alpha}$ where $A$ is the amplitude and $\alpha = 0.47 \pm 0.16$), in line with the tube-like scenario.

Our preliminary calculations above show that the magnetic flux tube shape can be determined conclusively if width, depth and energy of spectral lines are studied concurrently. Indeed, with these calculations we demonstrate that the loop shape differs from previously discussed and resembles a more complex 3D structure, while other loop shapes remain to be tested. This emphasizes the need in modelling for spectral line width and depth to obtain a more comprehensive picture of magnetar emission and geometry. 

\section{Future of Spectral Lines with New Detectors}
Motivated by our modelling results, we conducted several simulations to obtain a general view on magnetar spectral line detections with future X-ray telescopes. For all simulations, we created fake spectra using the \textit{fakeit} function on \textit{XSPEC} v12.11.1 (\cite[Arnaud 1996]{xspec})) with the most up-to-date public effective area and response of several detectors. For persistent emission, we represented the spectrum as an absorbed blackbody plus power law with a cylotron absorption feature as in Equation \ref{equation:cycl_model} (\textit{wabs $\times$ (bb+po) $\times$ cyclabs} on \textit{XSPEC}) setting all parameters to those reported in \cite[Tiengo et. al. (2013)]{tiengo2013} (best-fit model of the phase-averaged spectrum for the continuum model and reported parameters between phases 0$-$0.3 for the line model). For bursts, we used an absorbed Bremsstrahlung model with a Gaussian line (\textit{wabs $\times$ brems+gauss} on \textit{XSPEC}) with parameters set to the 6.4 keV line detection by \cite[Strohmayer et. al. (2000)]{strohmayer2000} during the 1998 burst precursor of SGR 1900+14, but scaled continuum normalization to a burst-like count rate of $\sim$ 35k in 0.2 seconds of burst duration. We then fit the simulated spectra with each detector's associated response and effective area, setting all parameters free with the exception of the persistent emission continuum model (which we fixed to the reported phase-averaged best-fits in \cite[Tiengo et. al. (2013)]{tiengo2013}) and SGR 1900+14 Hydrogen Column Density (fixed to the best-fit value in \cite[Strohmayer et. al. (2000)]{strohmayer2000}). 

\begin{figure}[htbp]
\centering
\xincludegraphics[trim={2.6cm 0.5cm 1.5cm 1cm}, clip,width=0.16\textwidth,label=a, angle=270]{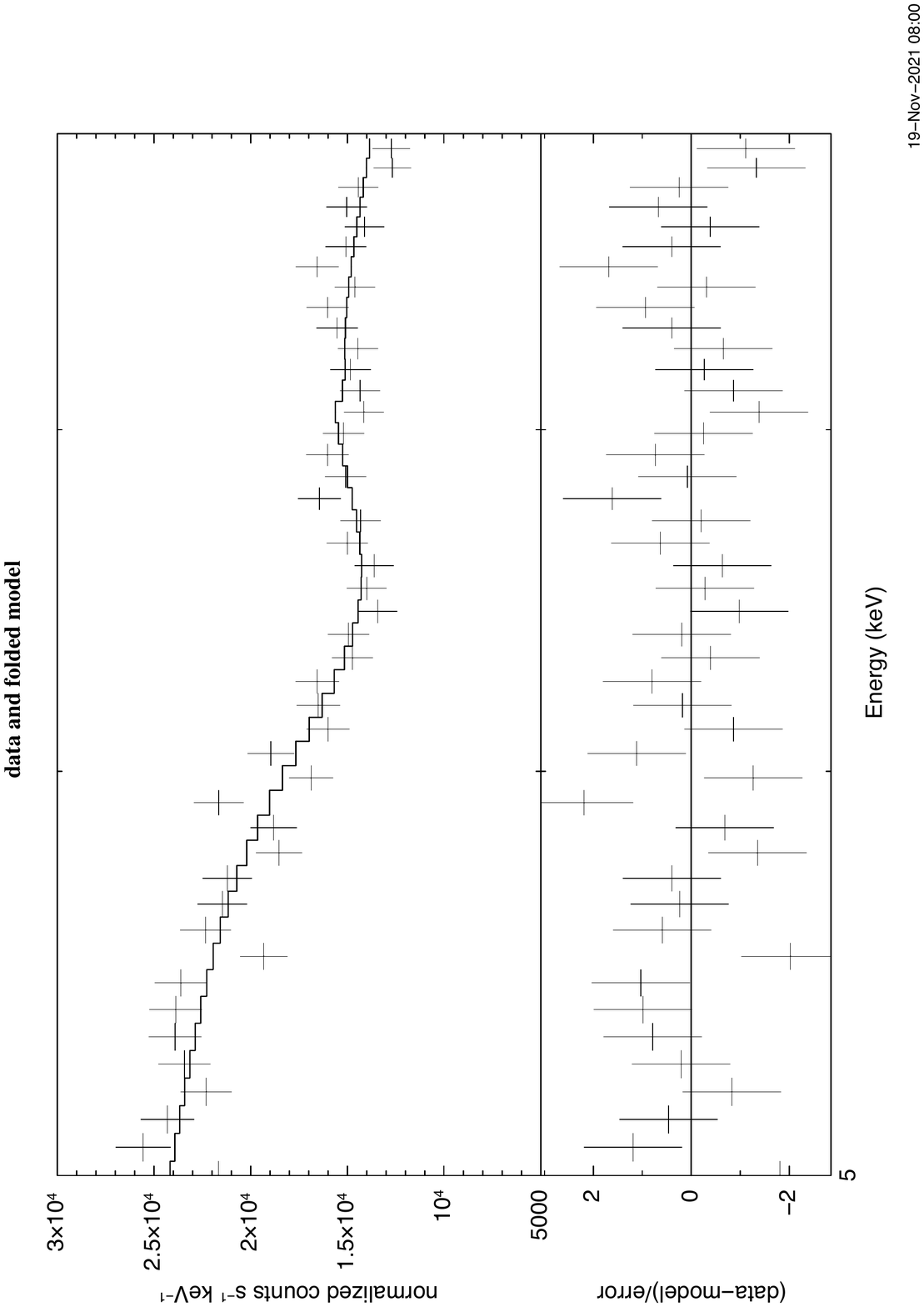}
\xincludegraphics[trim={2.6cm 0.5cm 1.5cm 1cm}, clip,width=0.16\textwidth,label=b, angle=270]{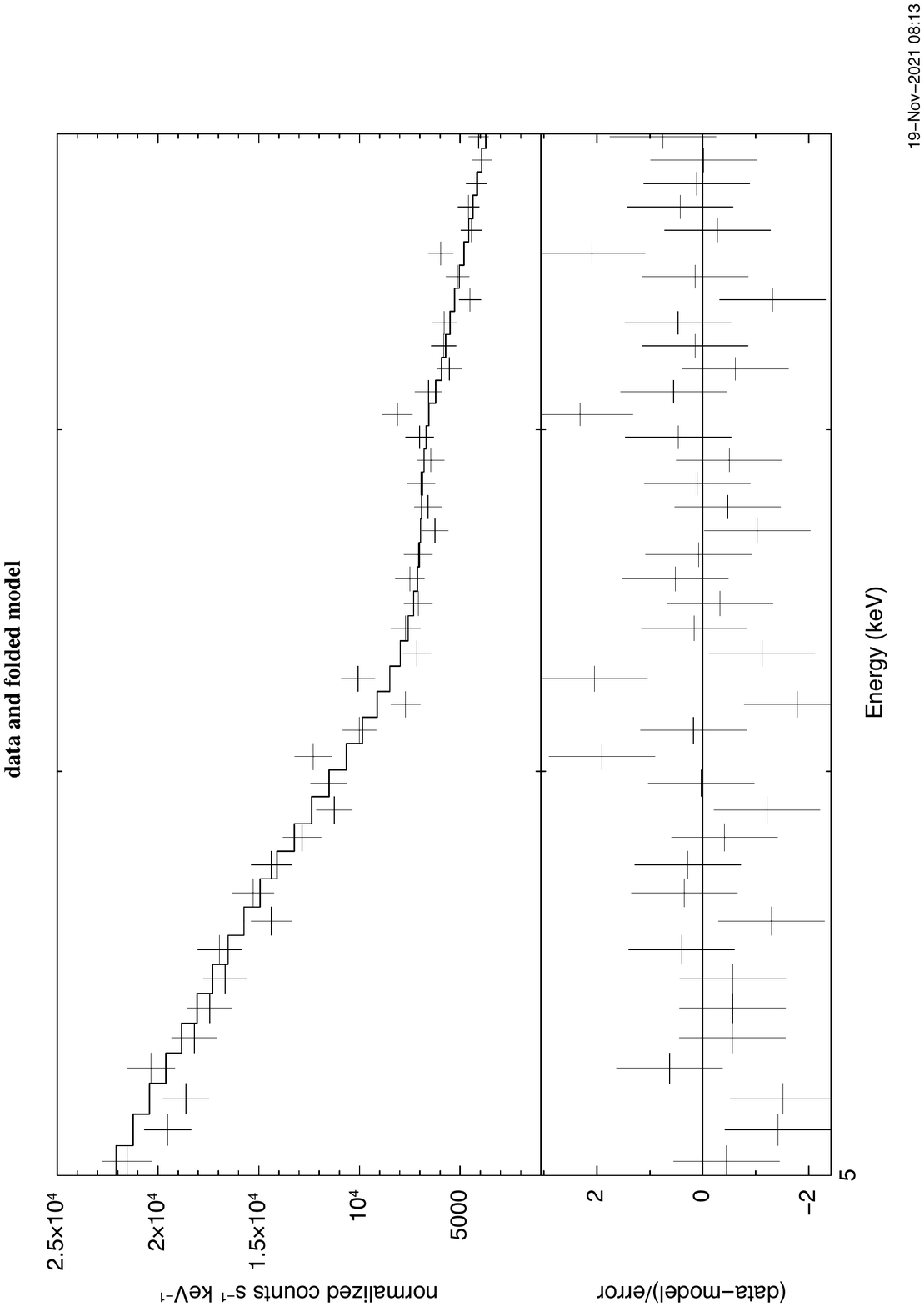}
\xincludegraphics[trim={2.6cm 0.5cm 1.5cm 1cm}, clip,width=0.16\textwidth,label=c, angle=270]{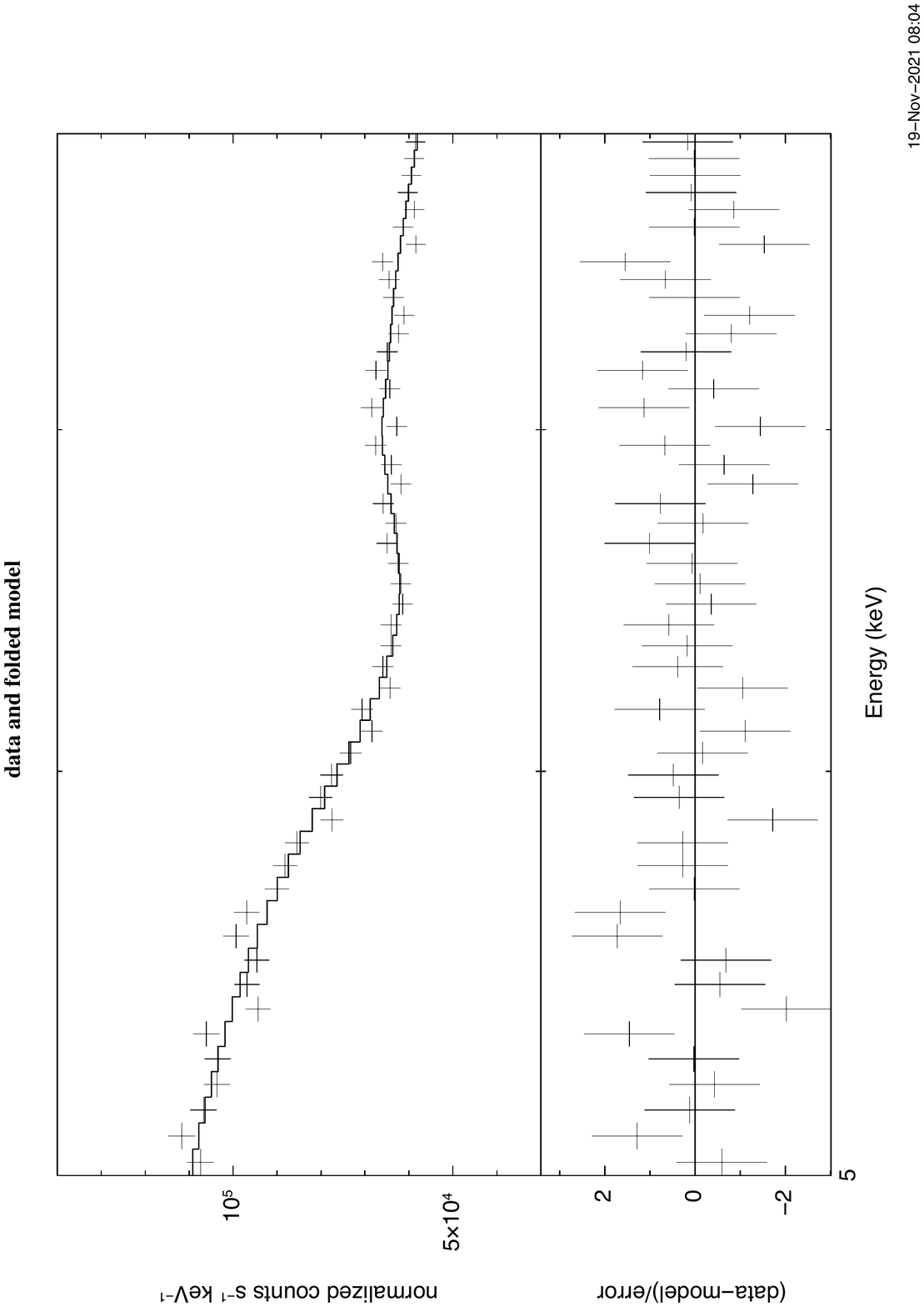} 
\xincludegraphics[trim={2.6cm 0.5cm 1.5cm 1cm}, clip,width=0.16\textwidth,label=d, angle=270]{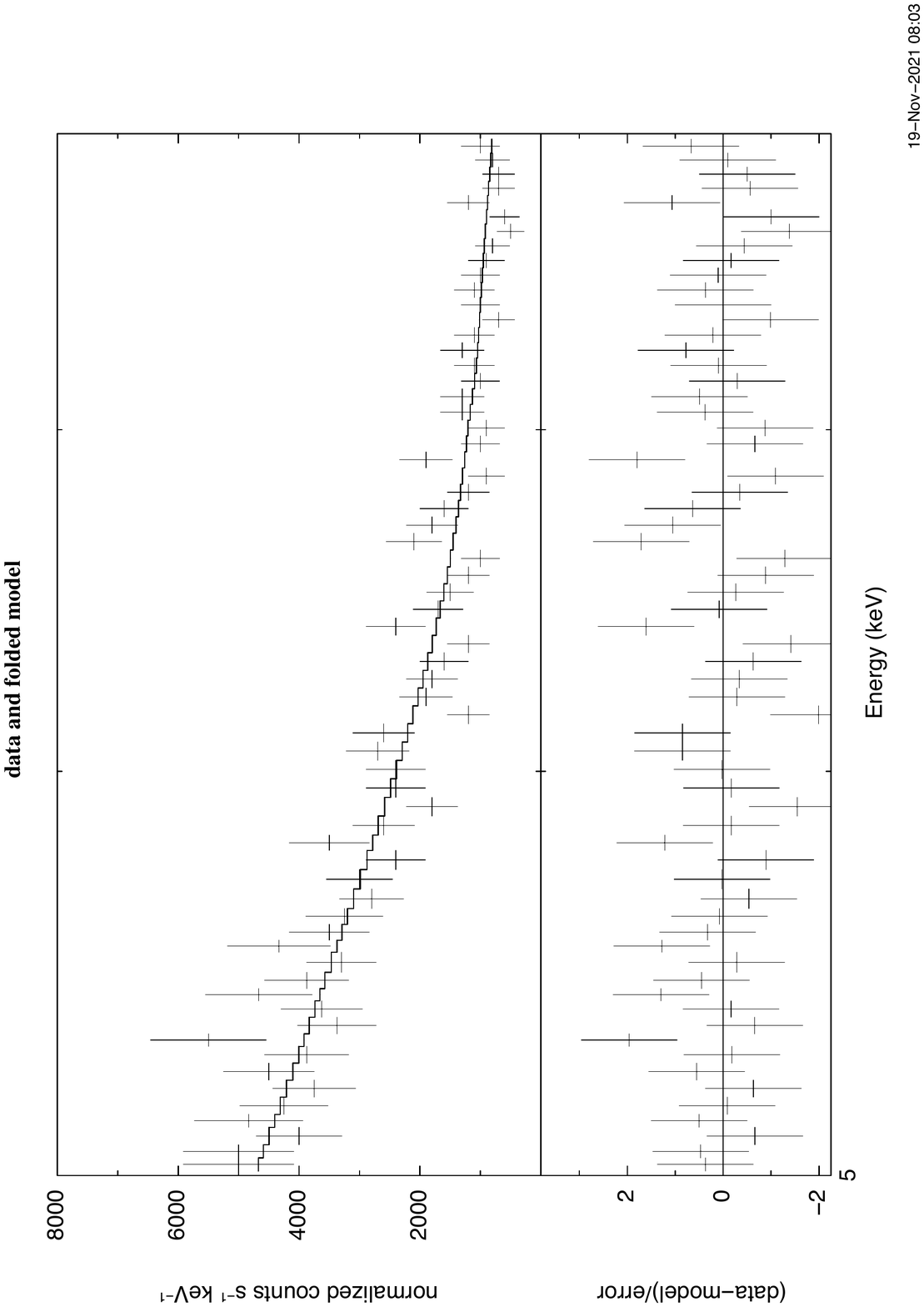}

\caption{SGR 1900+14 6.4 keV burst line fits and residuals with Colibr\`i (a) Athena X-IFU (b), eXTP LAD (c) and NICER (d) response and effective area.}
\label{fig:burstfits_all}
\end{figure}

For persistent emission of SGR 0418+5729, we plot average 1$\sigma$ line parameter errors per fitted value on the right panel of Figure \ref{fig:edw}. We find that line width, energy and depth are further constrained by over a factor of three with the advent of eXTP, Athena and Colibr\`i (see \cite[Heyl et. al. (2019)]{Heyl2019} for a description of Colibr\`i) compared to current detectors. The simulated burst lines and model fits for SGR 1900+14 in Figure \ref{fig:burstfits_all}, show that high effective area is crucial, with eXTP and Colibr\`i providing most accurate detections. The advent of large effective area telescopes with high pile-up limitations therefore place us on the edge of an exciting new era to probe magnetars with spectral lines.

\end{document}